\documentclass[english,times,11pt,a4paper]{paper}
\usepackage[T1]{fontenc}
\usepackage[latin9]{inputenc}
\usepackage{babel}
\usepackage{varioref}
\usepackage{url}
\usepackage{amsthm}
\usepackage{amsmath}
\usepackage{graphicx}
\usepackage[space]{grffile}

\makeatletter

\providecommand{\tabularnewline}{\\}
\newcommand{\lyxdot}{.}

\numberwithin{equation}{section}
\numberwithin{figure}{section}

\usepackage{subfig}
\makeatother

\begin{document}

\title{A Simulator for Data-Intensive Job Scheduling}

\author{Matteo Dell'Amico\\
EURECOM, Sophia Antipolis, France}
\maketitle

\begin{abstract}
Despite the fact that size-based schedulers can give excellent results
in terms of both average response times and fairness, data-intensive
computing execution engines generally do not employ size-based schedulers,
mainly because of the fact that job size is not known a priori.

In this work, we perform a simulation-based analysis of the performance
of size-based schedulers when they are employed with the workload
of typical data-intensive schedules and with approximated size estimations.
We show results that are very promising: even when size estimation
is very imprecise, response times of size-based schedulers can be
definitely smaller than those of simple scheduling techniques such
as processor sharing or FIFO.
\end{abstract}

\section{Introduction}

When scheduling batch jobs -- i.e., non-interactive programs -- the
main goal is to make sure that jobs are \emph{completed} as soon as
possible, as opposed to scheduling interactive processes, which should
progress at all time. For this reason, the so-called \emph{fair} scheduling
policies that divide evenly resources between running jobs are not
necessarily the most appropriate for batch jobs.

When the size of a job is known beforehand, \emph{size-based} policies
are effective. In fact, SRPT~\cite{schrage1966queue} is known to
obtain the minimum mean \emph{sojourn time} (i.e., the time that passes
between job submission and their completion) between all jobs; FSP~\cite{fsp}
provides a mean sojourn time close to the one of SRPT while preserving
fairness, in the sense that no jobs completes after the time they
would complete if using a ``fair'' processor sharing scheduling
discipline.

In this work, we study the applicability of size-based scheduling
in the field of big-data batch processing. There are two main peculiarities
that apply to such field, and the goal of this work is to evaluate
how they impact on the feasibility of implementing size-based scheduler
in such systems.
\begin{enumerate}
\item Job sizes vary by orders of magnitude~\cite{workloads,workloads_research}:
between a few seconds and several hours. This appears beneficial to
size-based scheduling solutions, since giving priority to smaller
jobs would entail huge benefits to them without impacting substantially
on the completion time of larger ones.
\item Job size is not perfectly known a priori. However, there are several
recent works that are able to \emph{estimate} job size~\cite{ARIA11,mascots12,nsdi12-c,query_perf}:
this approximate information can be used to inform scheduling. Of
course, when job size is estimated rather than known in advance, it
is impossible to guarantee minimality in all cases.
\end{enumerate}
Lu et al.~\cite{inaccuratesizebasedscheduling} provide results that
analyse experimentally the performance of size-based schedulers in
the presence of size estimation errors. However, those results are
not directly usable in our context, as inter-job arrival times and
job sizes are generated synthetically and they are not representative
of our use case; for this reason, the results of that work cannot
be used directly in our case. In addition, the FSP scheduler~\cite{fsp}
which is implemented both in the simulator of Lu et al. and in our
simulator has a degree of freedom when there are size estimation errors
(see Section \ref{sub:Implemented-Schedulers}); we show experimentally
that what could be considered as a minor implementation detail has
major effects on scheduling quality.

Given that the existing related work cannot give us a definite answer
to the question of how job size estimation errors could impact the
quality of scheduling in the context of big data batch system, we
built a custom simulator in order to evaluate that. The simulator,
described in detail in Section~\ref{sec:Simulator-Implementation},
performs a series of assumptions that abstracts away from the technicalities
and complexity of particular execution engines (such as, e.g., Hadoop,
Spark or Dryad), and we are using it to drive the design of the HFSP
Hadoop size-based scheduler~\cite{hfsp_arxiv}.

The simulation results shown in Section~\ref{sec:Simulation-Results}
allow us to conclude that size-based scheduling is very promising
for the field we are considering, since, in particular when the aging
technique is applied, it consistently and very significantly outperforms
both first-come-first-serve and fair-sharing schedulers.

\section{\label{sec:Simulator-Implementation}Simulator Implementation}

Our simulator is written in Python, and it requires the \texttt{numpy}
and \texttt{matplotlib} modules. It is available as free software.%
\footnote{\url{https://bitbucket.org/bigfootproject/schedsim}%
} In the following, we detail the assumptions that lead to our implementation
choices, and the way we parse existing Hadoop traces in order to assign
them to our simulator.

\subsection{Assumptions}

Schedulers for real-world data-intensive execution engines are complex,
since they have to consider a myriad of aspects related to the architectural
choices of the systems at hand. In this work, we take a simple approach
that abstracts away from them, reaping two benefits: the first one
is \emph{simplicity}, letting us define each job simply as an (arrival
time, execution time) pair and letting us implement traditional scheduling
policies exactly as they are defined in the literature; the second
one is \emph{generality}: our results are not influenced by the details
of a given execution engine. For system-related details, and their
evaluation on real workloads, we remand to our system work describing
the HFSP scheduler developed for Hadoop~\cite{hfsp_arxiv}, which
is currently the most widely used execution engine for data-intensive
systems.

In the following, we outline and motivate our assumptions.

\paragraph*{Resource Allocation}

Jobs are often divided in granular \emph{tasks}, and schedulers generally
have the duty to allocate those tasks to a discrete number of \emph{task
slots }available in the cluster. Two assumptions are related to resource
allocations.
\begin{enumerate}
\item The granularity of tasks is small enough that

\begin{enumerate}
\item whenever a job is preempted, its tasks can be considered to stop working
istantaneously;
\item the number of tasks per job is much larger than the number of task
slots, so that each job can run in parallel on the whole cluster.
\end{enumerate}

Using smaller tasks is actually advised in order to deal with the
problems of unfairness, stragglers and task size skew~\cite{tinytasks}.

\item The number of task slots is large enough that each job can be allocated
to run on an arbitrary fraction of the total system slots. This assumption
lets us implement perfect ``processor-sharing'' scheduling, running
each pending job on the same fraction of system resources.
\end{enumerate}

\paragraph*{Work Conservation}

We assume that the running time of a job's tasks is not influenced
by the time or choice of task slot it is run onto. In particular,
this means that each job will require the same amount of total resources,
without any penalty for having been preempted and resumed, disregarding
any data locality issues. We remark that architectures that avoid
penalties due to data locality have been proposed and successfully
implemented \cite{flat-datacenter-storage}.

\paragraph*{Error Distribution}

In this work, we consider log-normally distributed error values. In
particular, a job having size $s$ will be estimated as $\hat{s}=sX$,
where $X$ is a random variable with distribution $\operatorname{Log-\mathcal{N}}(0,\sigma^{2})$:
the choice of the log-normal distribution reflects the intuition that
an under-estimation $\hat{s}=s/k$ ($k>1$) is as likely as
an over-estimation $\hat{s}=ks$. When evaluating the performance
of the HFSP Hadoop scheduler on real jobs containing skew and stragglers,
we found that a log-normal distribution does indeed approximate well
the empirically observed values for estimation error in our case.

\subsection{Parsing SWIM \texttt{.tsv} files}

SWIM~\cite{swim_tool} is a well-known tool to generate workloads
to test MapReduce systems; it has been used in academia to validate
proposals to improve Hadoop (see e.g.~\cite{eurosys10,swim}). SWIM
ships with samples of traces from Facebook: for each job $j$ in those
traces, they contain:
\begin{enumerate}
\item Job submission time $t_{j}$;
\item Input size (from disk) $i_{j}$;
\item Size of data ``shuffled'' on the network $s_{j}$;
\item Output size (to disk) $o_{j}$.
\end{enumerate}
We combine points 2--4 in a single value, representing the number
of seconds that the system would need to execute these jobs if they
were running using all the cluster resources. If the whole system
can read and write data from disk at speed $d$ and send it over the
network at speed $n$, we consider the size of job $j$ as
\[
S_{j}=d\left(i_{j}+o_{j}\right)+ns_{j}.
\]

In our system, rather than specifying $d$ and $n$, we want however
to evaluate scheduler performance based on a more abstract notion
of \emph{load. }We prefer, therefore, to characterize our system as
heavily or lightly loaded, and having a given disk / network performance
ratio. We do so by fixing the ratio $d/n$ that represents the ratio
between the aggregate disk and network bandwidth of the whole system
(a value of 1 would represent a system where the network is never
the bottleneck such as Flat Datacenter Storage~\cite{flat-datacenter-storage},
while a higher value is representative of more traditional installations)
and a load value $l$ that represents the ratio between the total
size of all jobs and the time passing between the instant $t_{0}$
of submitting the first job and $t_{e}$, when the last job is submitted.
We obtain the values $d$ and $n$, and therefore the value $S_{j}$
for the size of each job, by solving the following set of equations:
\[
\begin{cases}
\sum_{j}S_{j}=\sum_{j}d\left(i_{j}+o_{j}\right)+ns_{j}=l\left(t_{e}-t_{0}\right)\\
d/n=X,
\end{cases}
\]

where $X$ is a user-set value. In the following of the paper, we
use default values of $l=0.9$ and $d/n=4$, to account for highly
loaded systems with more disk bandwidth than network bandwidth. Table~\vref{tab:Simulator-parameters}
summarizes the system parameters.

\begin{table*}
\begin{centering}
\begin{tabular}{|c|c|c|}
\hline 
Name & Default & Meaning\tabularnewline
\hline 
\hline 
$d/n$ & 4 & Ratio between disk and network bandwidth in the system\tabularnewline
\hline 
$l$ & 0.9 & Average load in the system\tabularnewline
\hline 
$\sigma$ & -- & Value for error distribution\tabularnewline
\hline 
\end{tabular}
\par\end{centering}

\caption{\label{tab:Simulator-parameters}Simulator parameters}
\end{table*}

\subsection{\label{sub:Implemented-Schedulers}Implemented Schedulers}

We implemented four schedulers: FIFO (First In First Out) and PS (Processor
Sharing) are traditional schedulers that do not need size estimation;
as size-based schedulers, we implemented SRPT (Shortest Remaining
Processing Time) and FSP (Fair Sojourn Protocol).

\paragraph*{FIFO}

This basic scheduling discipline is often also known as FCFS (First
Come First Serve). In it, jobs are scheduled the whole resources of
the system in the order of their arrival time. FIFO is known to perform
poorly in workloads where jobs of mixed sizes appear: our experimental
results confirm this, showing that FIFO is the worst-performing scheduling
discipline among those implemented.

\paragraph*{PS}

This technique is the considered a ``fair'' scheduling disciplines:
when there are $n$ pending jobs, each of them is allocated $1/n$-th
of the system resources. While this guarantees that all pending jobs
progress, none of them progresses quickly. As a result, in loaded
systems PS tends to result in many scheduled processes, each of them
progressing slowly.

\paragraph*{LAS}

Least Attained Service (LAS) is a scheduling discipline that allocates
resources to the job that had received the least service time. It is
insteresting to compare LAS to other disciplines and in our case,
since it favours small jobs and performs well in cases of skewed job
size distributions~\cite{Rai:2003:ALS:885651.781055}; unlike
size-based scheduling policies, however, it does not require knowledge
of job size.

\paragraph*{SRPT}

This technique, in absence of size estimation errors, minimize the
metric of mean \emph{sojourn time~\cite{schrage1966queue}} -- i.e.,
the time that passes between a job's submission and its completion.
It does so by assigning all system resources to the pending job that
requires the least remaining amount of work to complete, therefore
minimizing the number of pending jobs at each moment. SRPT differs
from SJF (Shortest Job First) in that the arrival of a new job having
size smaller than the remaining amount of work of a running one will
preempt the running one.

While SRPT optimizes mean sojourn time, it may not be fair, since
large running jobs may be denied access to resources for long if smaller
jobs are constantly submitted. In realistic use cases and in the absence
of errors, however, this phenomenon is known to be unlikely~\cite{moseley_soda}.

\paragraph*{FSP}

This scheduling discipline, proposed by Friedman and Henderson~\cite{fsp},
combines the fairness guarantees of PS with the performance improvements
obtained through size-based scheduling. It is similar in concept to
SRPT, but priority is given to jobs with the smallest remaining processing
time in a \emph{virtual} emulated system which is running PS. The
solution of virtually decreasing the size of jobs even when they are
not scheduled is called \emph{job aging}, and it avoids the starvation
that could happen in SRPT. In particular, the aging applied by FSP
guarantees fairness in the sense that (in the absence of size estimation
errors) jobs in FSP are guaranteed to complete \emph{not later} than
in PS. The same mechanism of FSP has been also proposed under the
name of \emph{fair queuing~}\cite{fair_queuing} and \emph{Vifi}~\cite{gorinsky2007fair}.

When considering size estimation errors, the definition of FSP gives
a degree of freedom to the implementation: what to do when one or
more pending job are ``late'', i.e. they reach a virtual size of
zero? The fairness properties of FSP guarantee that this will never
happen if there are no size estimation errors; however, when job size
is underestimated, this is a rather common event. In this case, we
implemented two alternative policies: 
\begin{itemize}
\item \textbf{FSP+FIFO}, which schedules late job according to a FIFO policy:
late jobs have priority over all other pending jobs, and the first
one to reach a virtual size of zero obtains all system resources;
\item \textbf{FSP+PS}, which shares equally the system resources between
late jobs: they have priority over all other pending jobs and each
of the $n$ late jobs get $1/n$-th of the system resources.
\end{itemize}
Our experimental results, shown in the following section, highlight
how this appearingly minor detail has major effects on the performance
of the scheduler.

\section{\label{sec:Simulation-Results}Simulation Results}

After describing the implementation of our simulator, we are now ready
to show our simulation results on the three workloads made available
with the SWIM tool~\cite{workloads}:
\begin{itemize}
\item \texttt{FB09-0}: a trace from Facebook in 2009, containing 5,894 jobs.
\item \texttt{FB09-1}: again a trace from Facebook in 2009, containing 6,638
jobs.
\item \texttt{FB10}: a 2010 trace with 24,442 jobs.
\end{itemize}
All results shown in this section are obtained by running 100 simulation
runs for each combination of input file, values of $\sigma$, and
settings for $l$ and $d/n$. Since, for given values of $l$ and
$d/n$, the trace is fixed, what changes between simulation runs are
only estimation errors. Therefore, multiple simulation runs are not
needed when there is no size estimation errors and for the FIFO and
PS schedulers.

\subsection{\label{sub:sigma}Sojourn versus $\sigma$}

\begin{figure*}
\begin{centering}
\subfloat[SRPT.]{\begin{centering}
\includegraphics[width=0.32\linewidth]{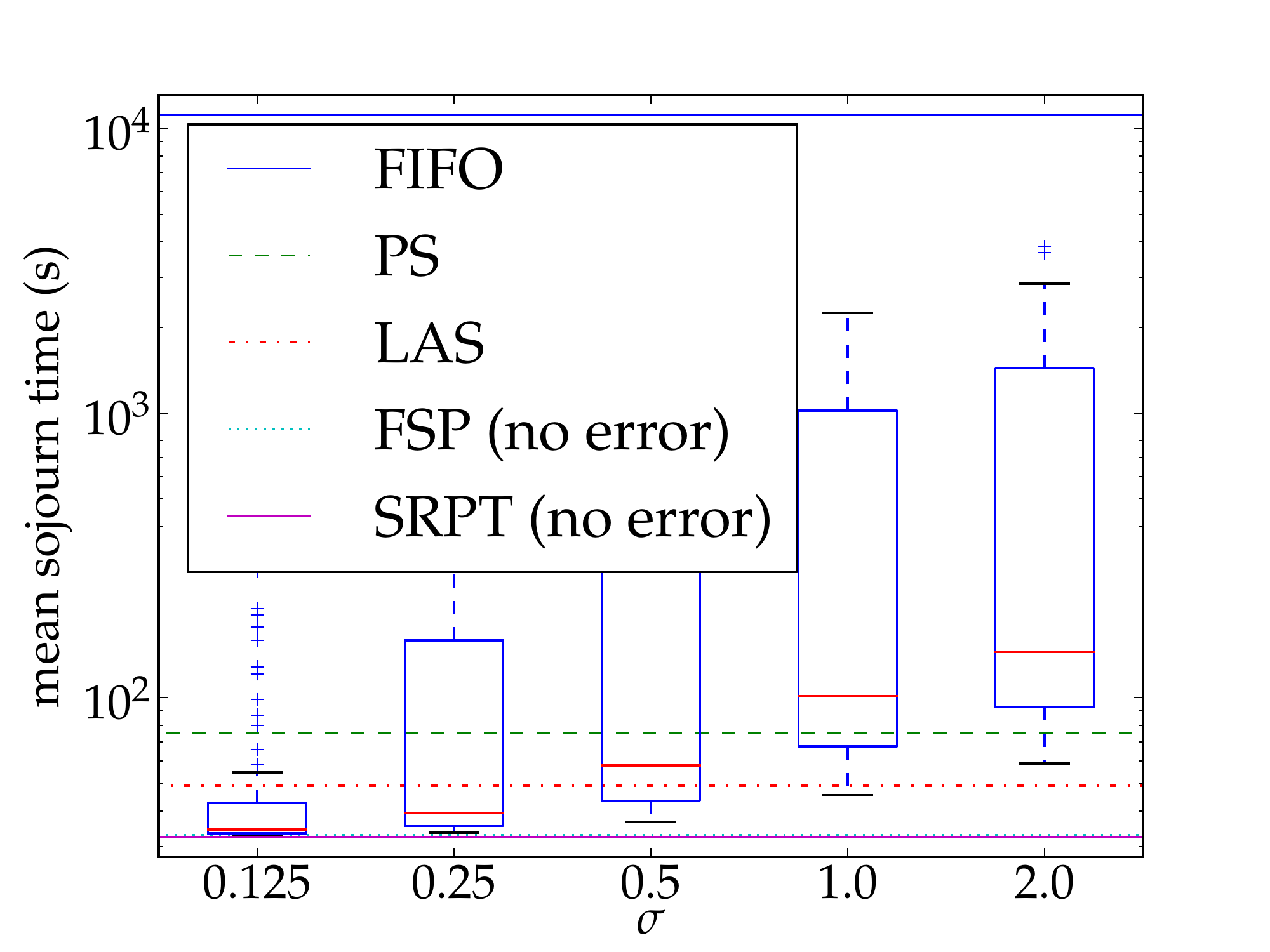}
\par\end{centering}

}\subfloat[FSP+FIFO.]{\begin{centering}
\includegraphics[width=0.32\linewidth]{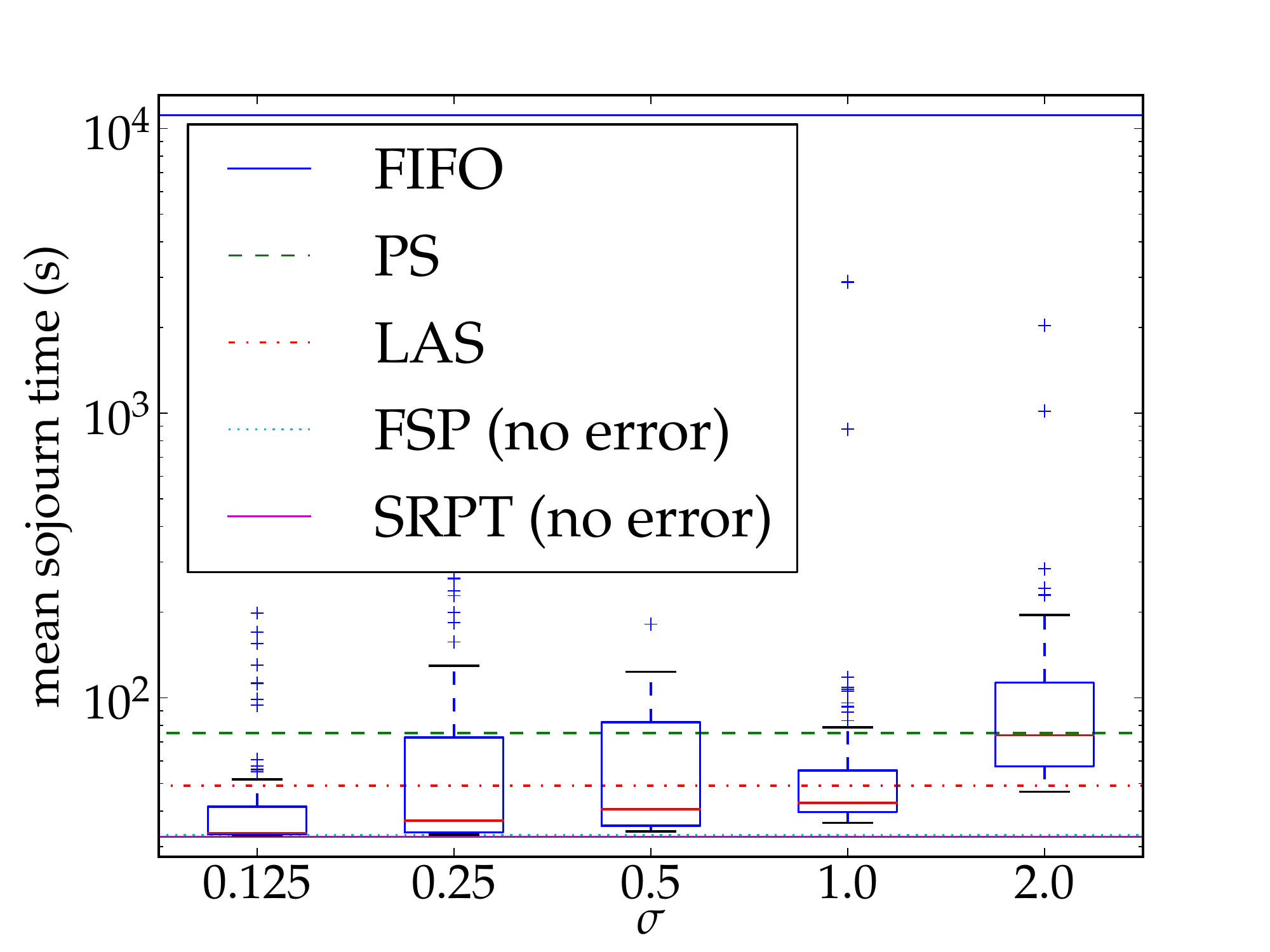}
\par\end{centering}

}\subfloat[FSP+PS.]{\begin{centering}
\includegraphics[width=0.32\linewidth]{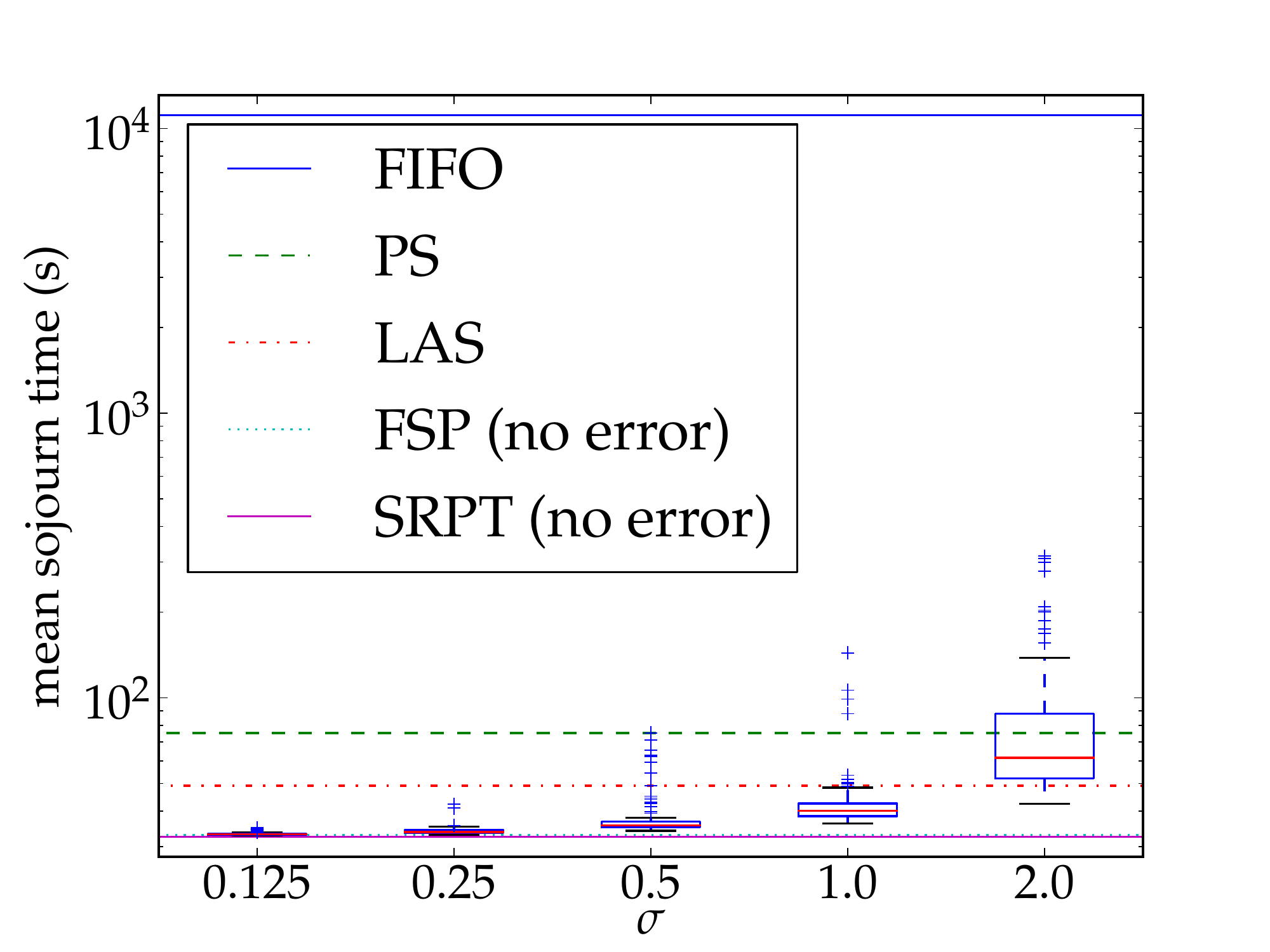}
\par\end{centering}

}\caption{\label{fig:sigma-fb09-0}Sojourn versus $\sigma$ on the \texttt{FB09-0}
workload.}

\par\end{centering}

\end{figure*}

\begin{figure*}
\centering{}\subfloat[SRPT.]{\begin{centering}
\includegraphics[width=0.32\linewidth]{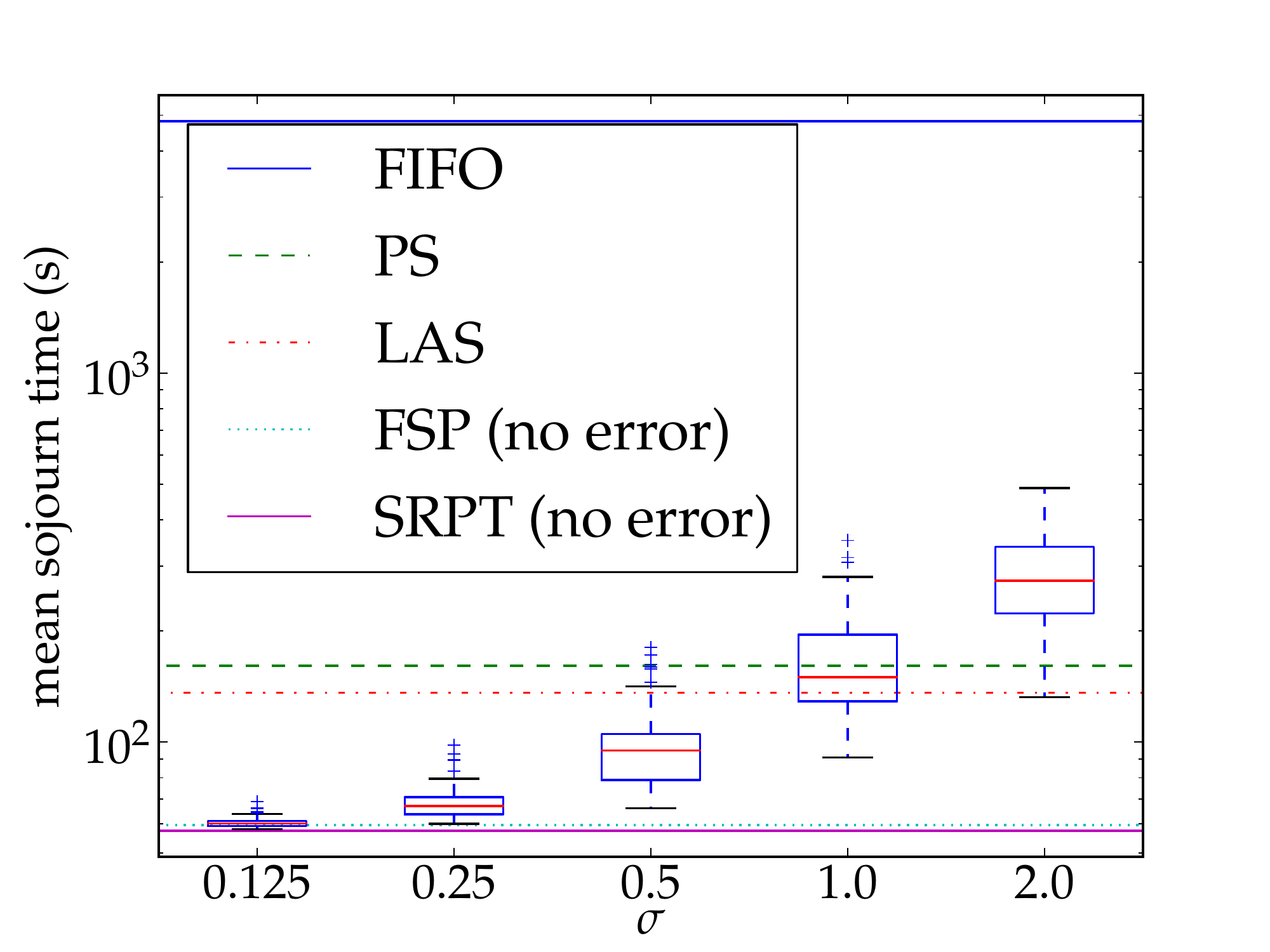}
\par\end{centering}

}\subfloat[FSP+FIFO.]{\begin{centering}
\includegraphics[width=0.32\linewidth]{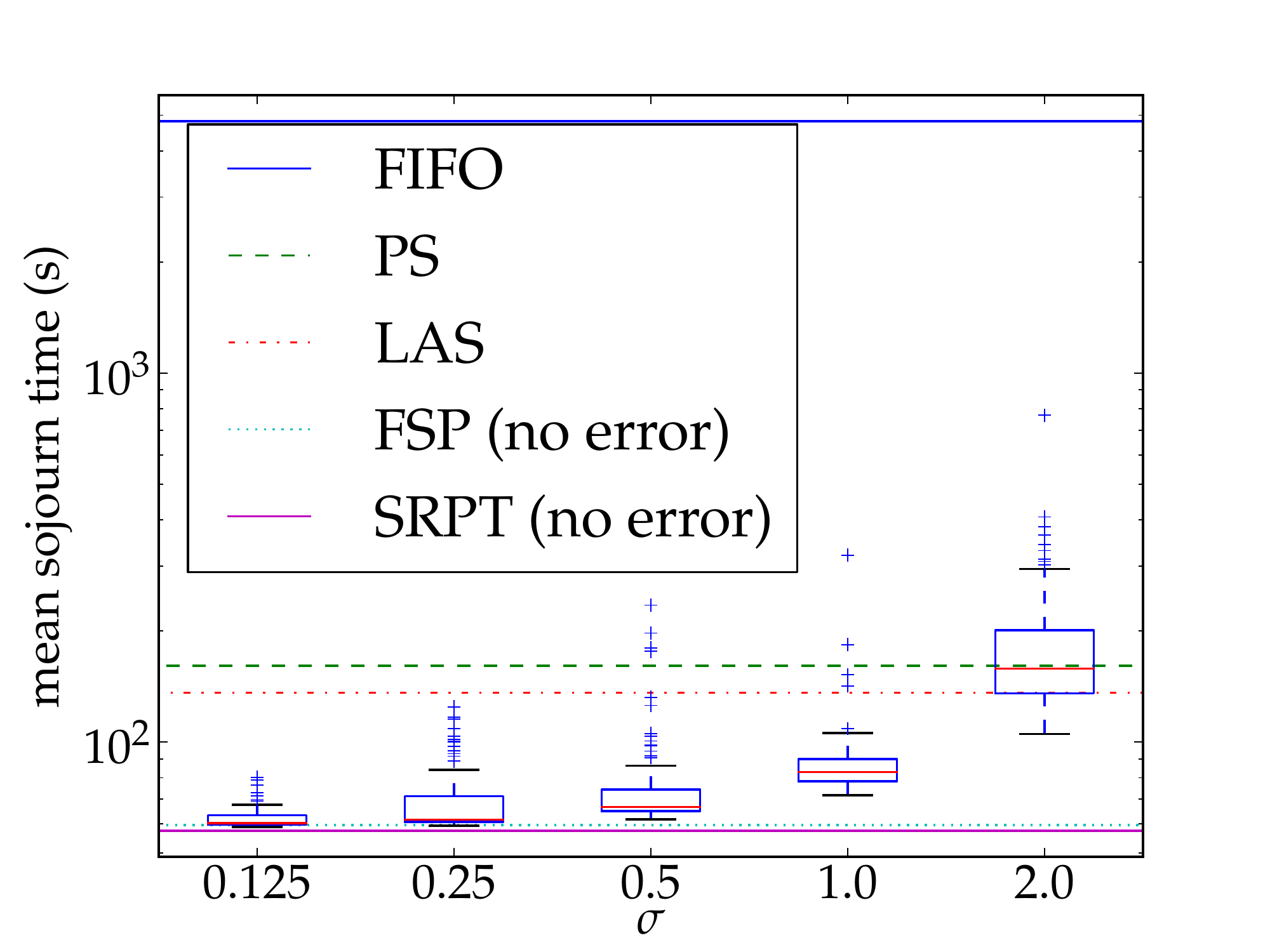}
\par\end{centering}

}\subfloat[FSP+PS.]{\begin{centering}
\includegraphics[width=0.32\linewidth]{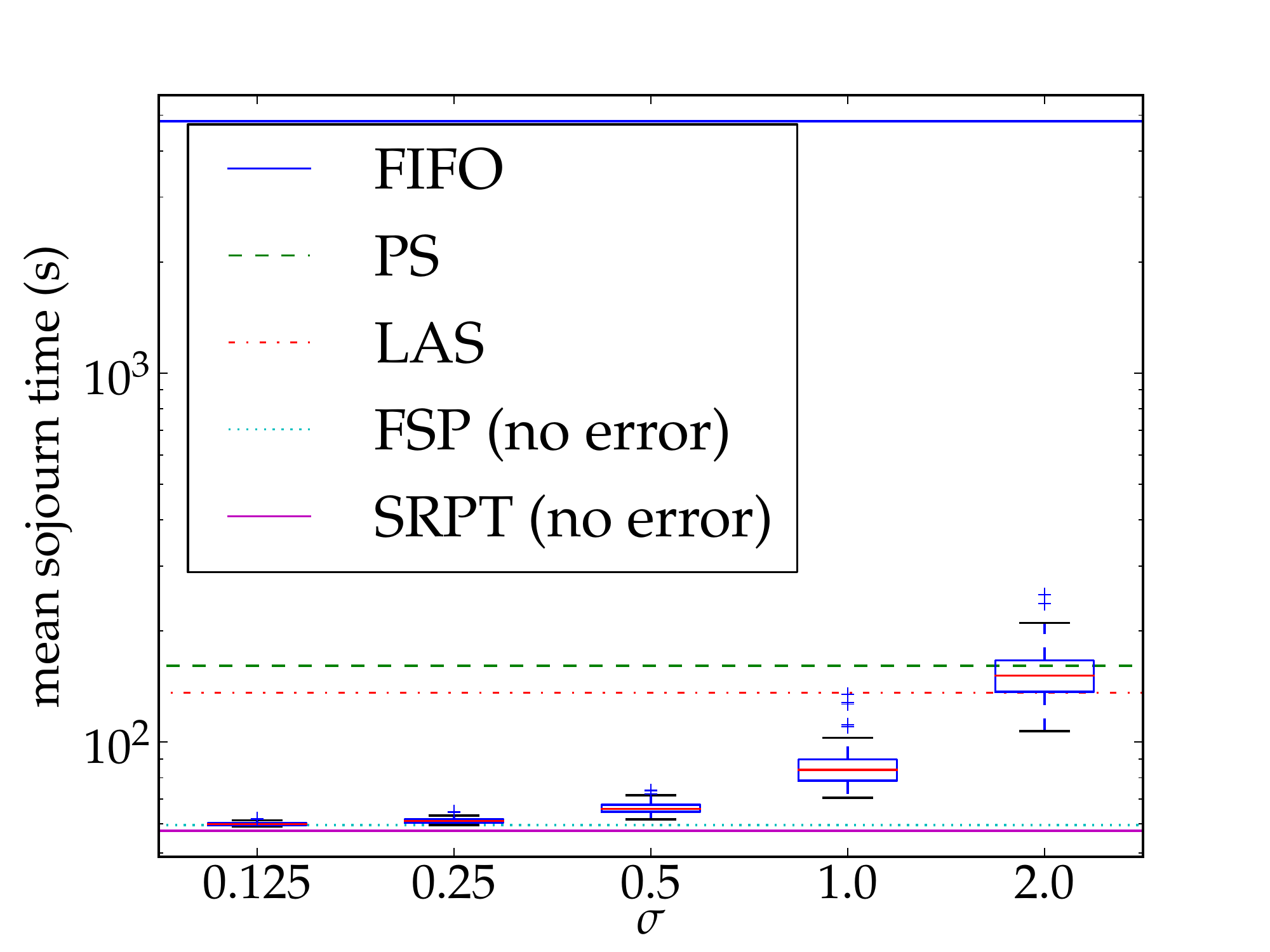}
\par\end{centering}

}\caption{\label{fig:sigma-fb09-1}Sojourn versus $\sigma$ on the \texttt{FB09-1}
workload.}
\end{figure*}

\begin{figure*}
\centering{}\subfloat[SRPT.]{\begin{centering}
\includegraphics[width=0.32\linewidth]{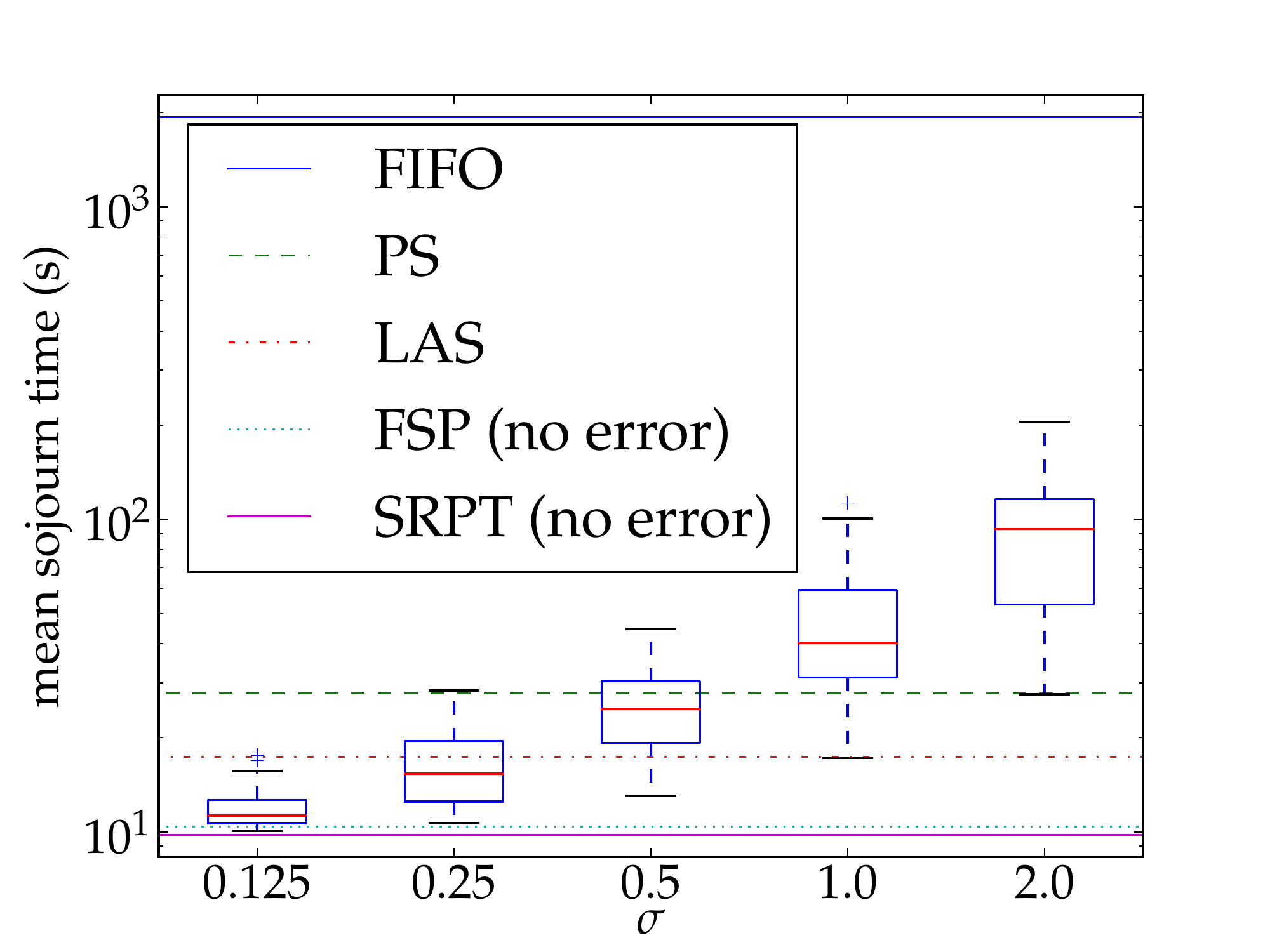}
\par\end{centering}

}\subfloat[FSP+FIFO.]{\begin{centering}
\includegraphics[width=0.32\linewidth]{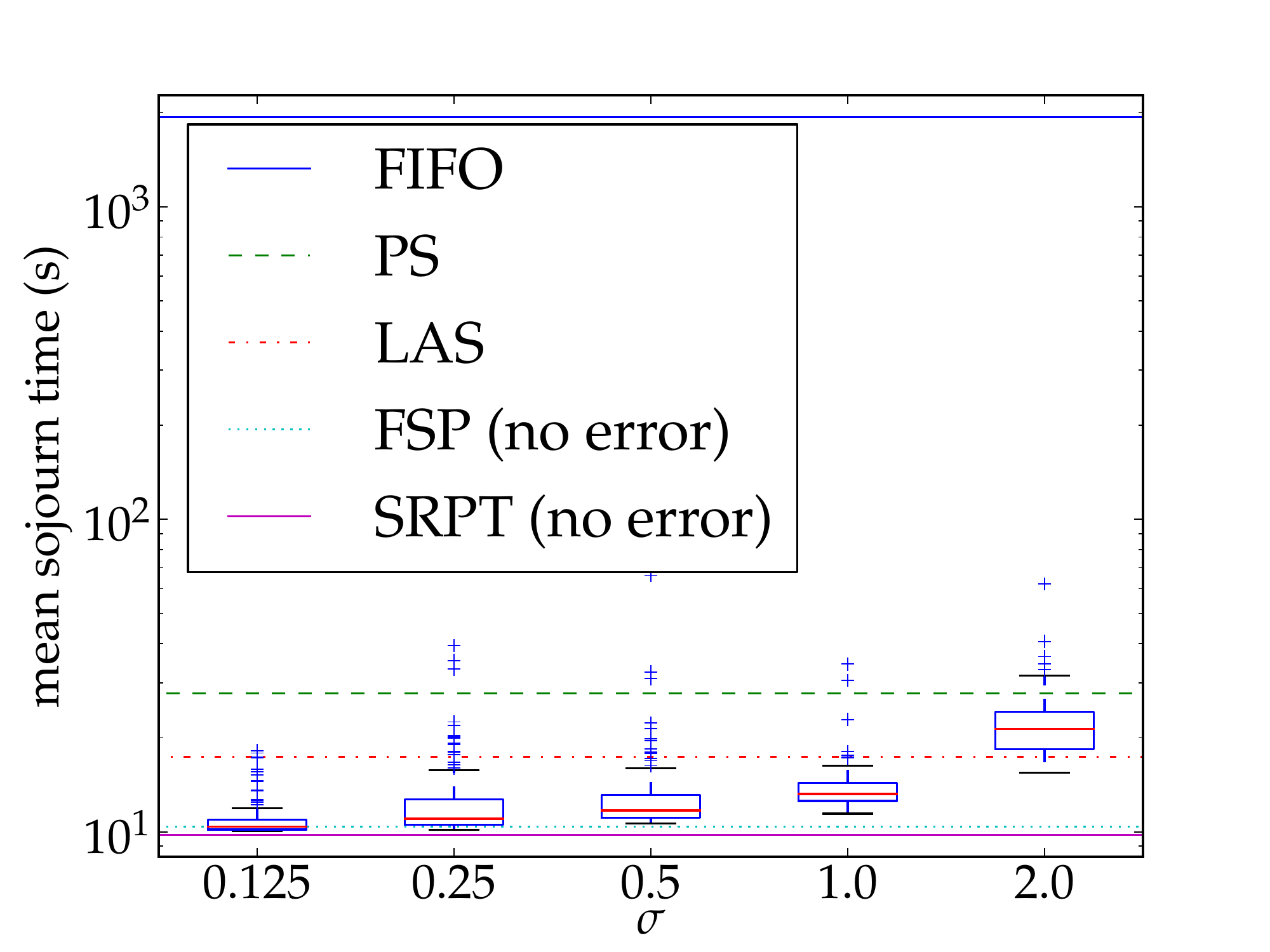}
\par\end{centering}

}\subfloat[FSP+PS.]{\begin{centering}
\includegraphics[width=0.32\linewidth]{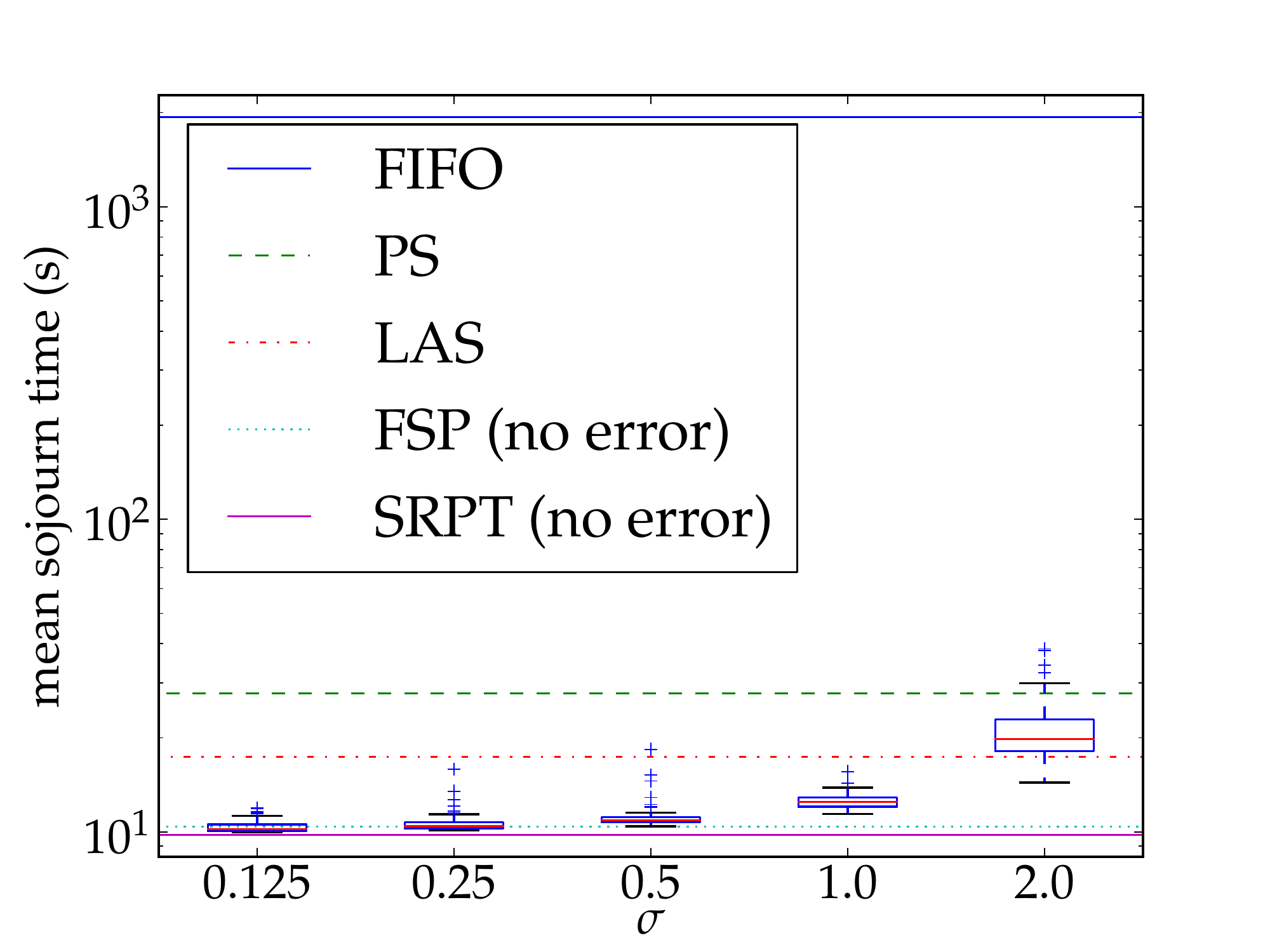}
\par\end{centering}

}\caption{\label{fig:sigma-fb10}Sojourn versus $\sigma$ on the \texttt{FB10}
workload.}
\end{figure*}

We start by investigating the impact of the $\sigma$ value which
describes the magnitude of errors, on mean sojourn time. Figures \ref{fig:sigma-fb09-0},
\ref{fig:sigma-fb09-1}, and \vref{fig:sigma-fb10} show a box-plot
(highlighting the median and the most important percentiles) for mean
sojourn times over the 100 experiment runs, for varying values of
$\sigma$. Since sojourn times vary by orders of magnitude, here and
in the following of the sections, they are plotted on a logarithmic
scale.

We can at first see that the FIFO scheduler, in this case where job
sizes differ by orders of magnitude, performs much worse than all
other scheduling primitives: therefore, it can be regarded as essentially
a worst case. By guaranteeing that each pending job progresses, PS
results in a sojourn time which is \emph{orders of magnitude }better.
For this reason, we consider the performance of PS as an ``acceptable''
one, and good performance whatever is able to outperform PS.

It is interesting to examine the performance of the LAS scheduler: it
is one that favours small jobs in situations, like ours, where job
size is heavily skewed. We can see that LAS generally performs better
than PS, and it is therefore a good candidate in cases like ours, when
job size is impossible to estimate, or it can only be estimated with
high error.

In accordance with intuition, we see that increasing the error rate
is detrimental to the performance of size-based schedulers. However,
SRPT does not handle errors terribly well, when compared to FSP. We
consider this is due to the fact that even large estimation errors
are, in the long run, corrected by aging: this avoids that even widely
over-estimated jobs are scheduled very late. In addition, we observe
that there is a notable difference in terms of performance between
FSP+FIFO -- which exhibits a few ``outlier'' experiment runs where
mean sojourn time is much higher -- and FSP+PS, where performance
is consistent between experiment runs. We explain this with the fact
that severe underestimation errors can result in long jobs being scheduled
too early in both cases, but while this does not produces catastrophic
effects in FSP+PS, where all ``late'' jobs progress, in FSP+FIFO,
even ``late'' jobs may do not progress for relevant amounts of time.
We conclude that FSP+PS is the best performing scheduling strategy
between those examined in the case of errors.

What is perhaps most surprising from these results is actually the
robustness of size-based schedulers, and in particular of FSP+PS,
to size estimation errors: even when $\sigma=1$, where in around
half of the cases there is an over- or under-estimation by a factor
of 2 or more, FSP+PS consistently and significantly outperforms the
PS scheduler. This lets us conclude that, according to the traces
we have at hand, size-based scheduling, and in particular FSP+PS,
appear \emph{very resilient} to estimation errors.

\subsection{Sojourn versus load}

We now turn our attention to the performance of scheduler when varying
load. In this case, we plot the average of mean sojourn time between
experiment intervals (we do not plot box-plots or confidence intervals
for readability), and we vary the $l$ parameter between 0.1 and 2.

\begin{figure*}
\centering{}\subfloat[\texttt{FB09-0}.]{\begin{centering}
\includegraphics[width=0.32\linewidth]{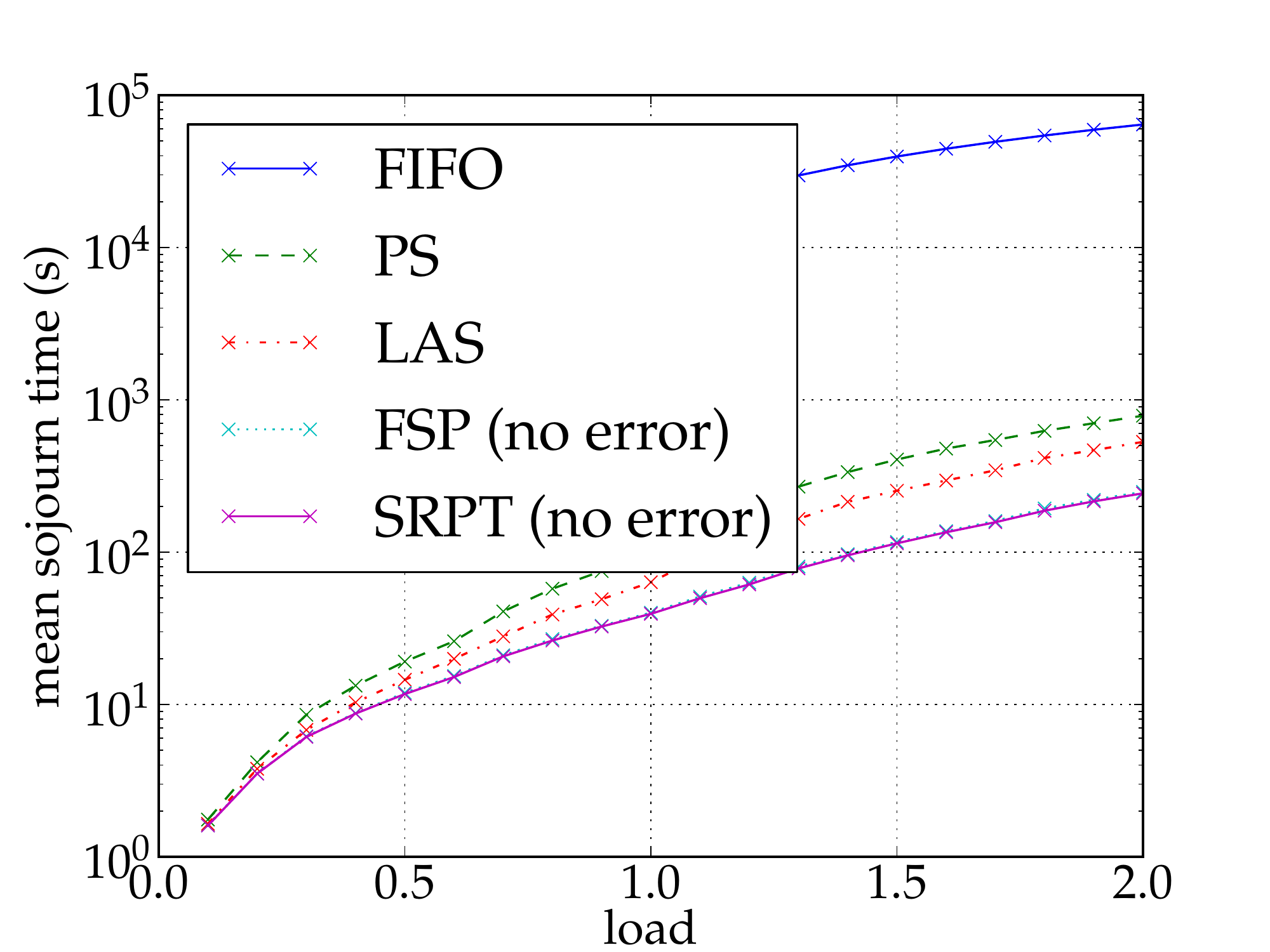}
\par\end{centering}

}\subfloat[\texttt{FB09-1}.]{\begin{centering}
\includegraphics[width=0.32\linewidth]{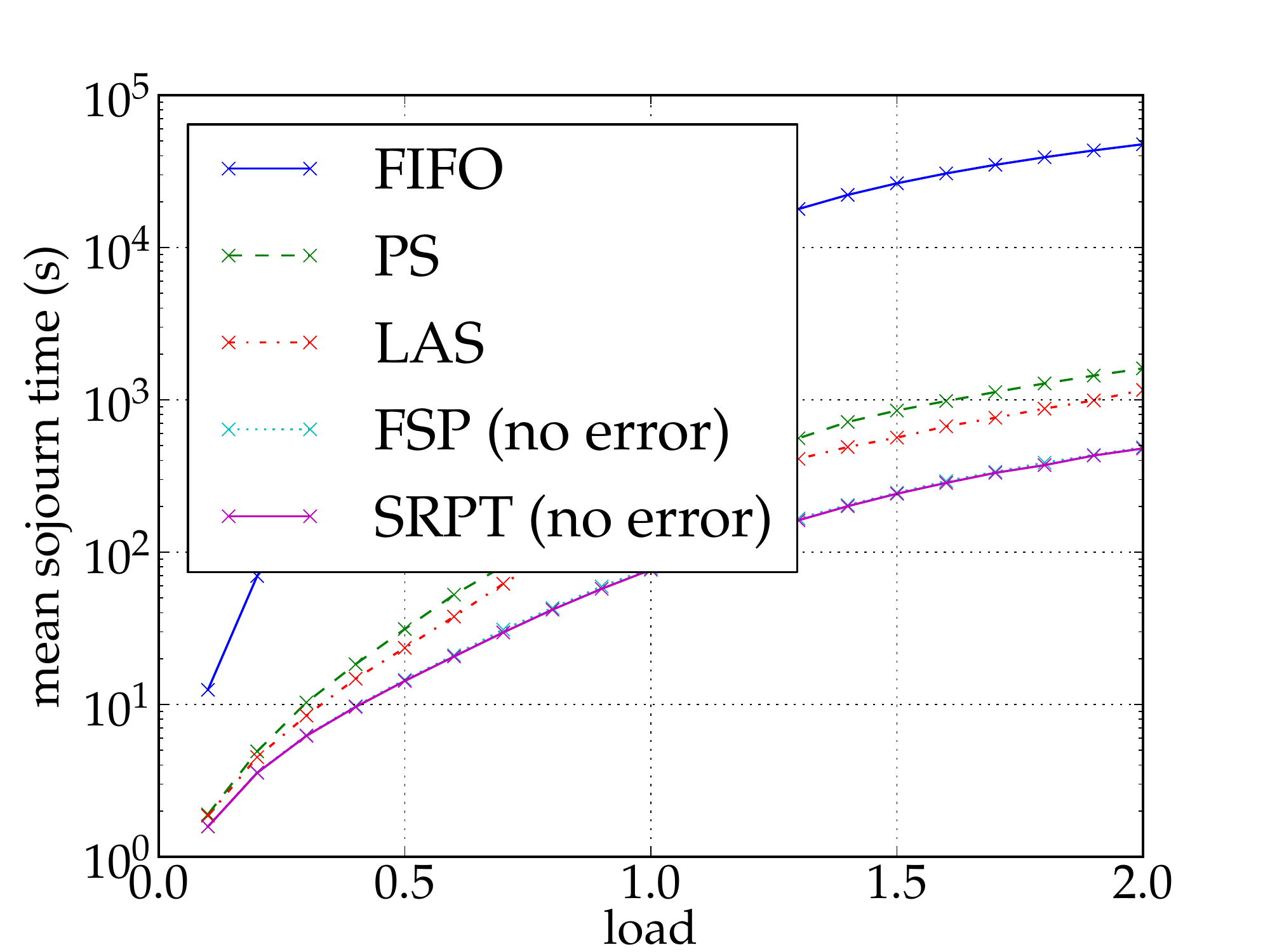}
\par\end{centering}

}\subfloat[\texttt{FB10}.]{\begin{centering}
\includegraphics[width=0.32\linewidth]{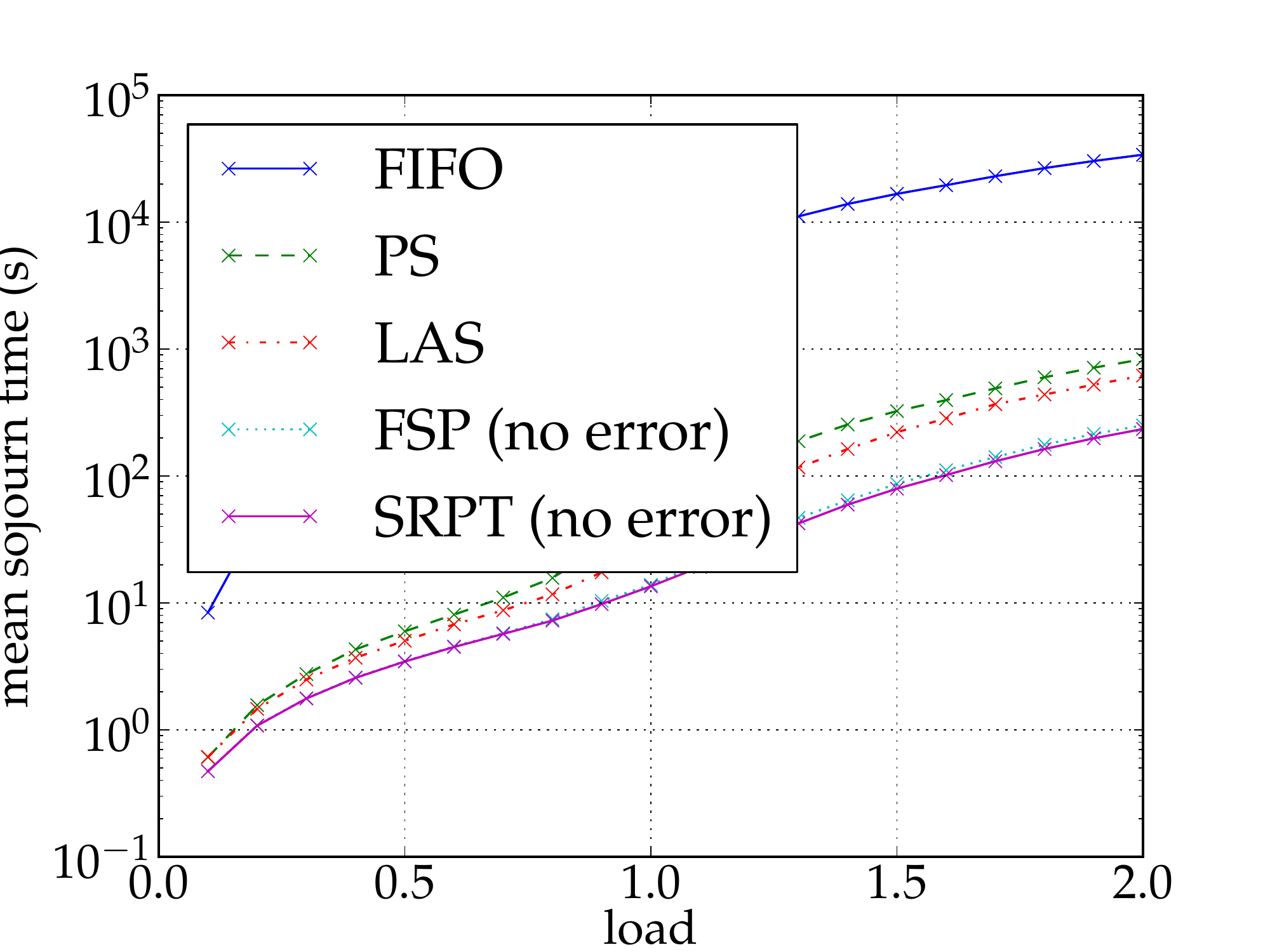}
\par\end{centering}

}\caption{\label{fig:load-0}Sojourn versus load: no error.}
\end{figure*}

In Figure \vref{fig:load-0}, we show how mean sojourn time increases
when increasing the load in the absence of size estimation errors:
we can see that sojourn time increases smoothly as load grows in all
the three datasets that we consider. Again, we confirm that FIFO can
be considered a worst case, with a mean sojourn time which is orders
of magnitude longer in all cases. We can also notice that FSP and
SRPT perfom in a remarkably similar way: even when there are no size
estimation errors, FSP's fairness guarantee comes at what appears
to be a negligible cost in term of mean sojourn time. These results
confirm those obtained by Friedman and Henderson~\cite{fsp}.

\begin{figure*}
\centering{}\subfloat[\texttt{FB09-0}.]{\begin{centering}
\includegraphics[width=0.32\linewidth]{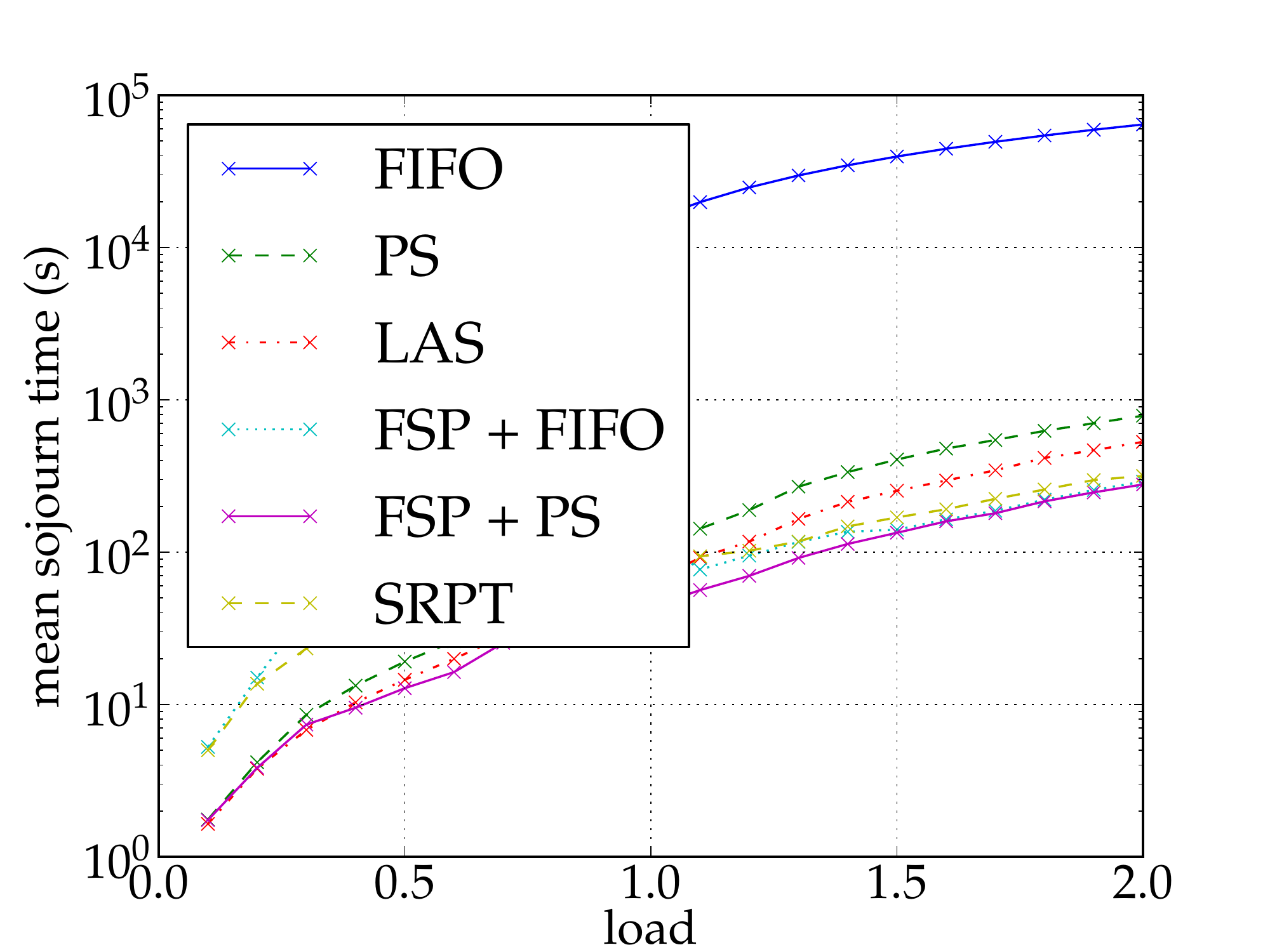}
\par\end{centering}

}\subfloat[\texttt{FB09-1}.]{\begin{centering}
\includegraphics[width=0.32\linewidth]{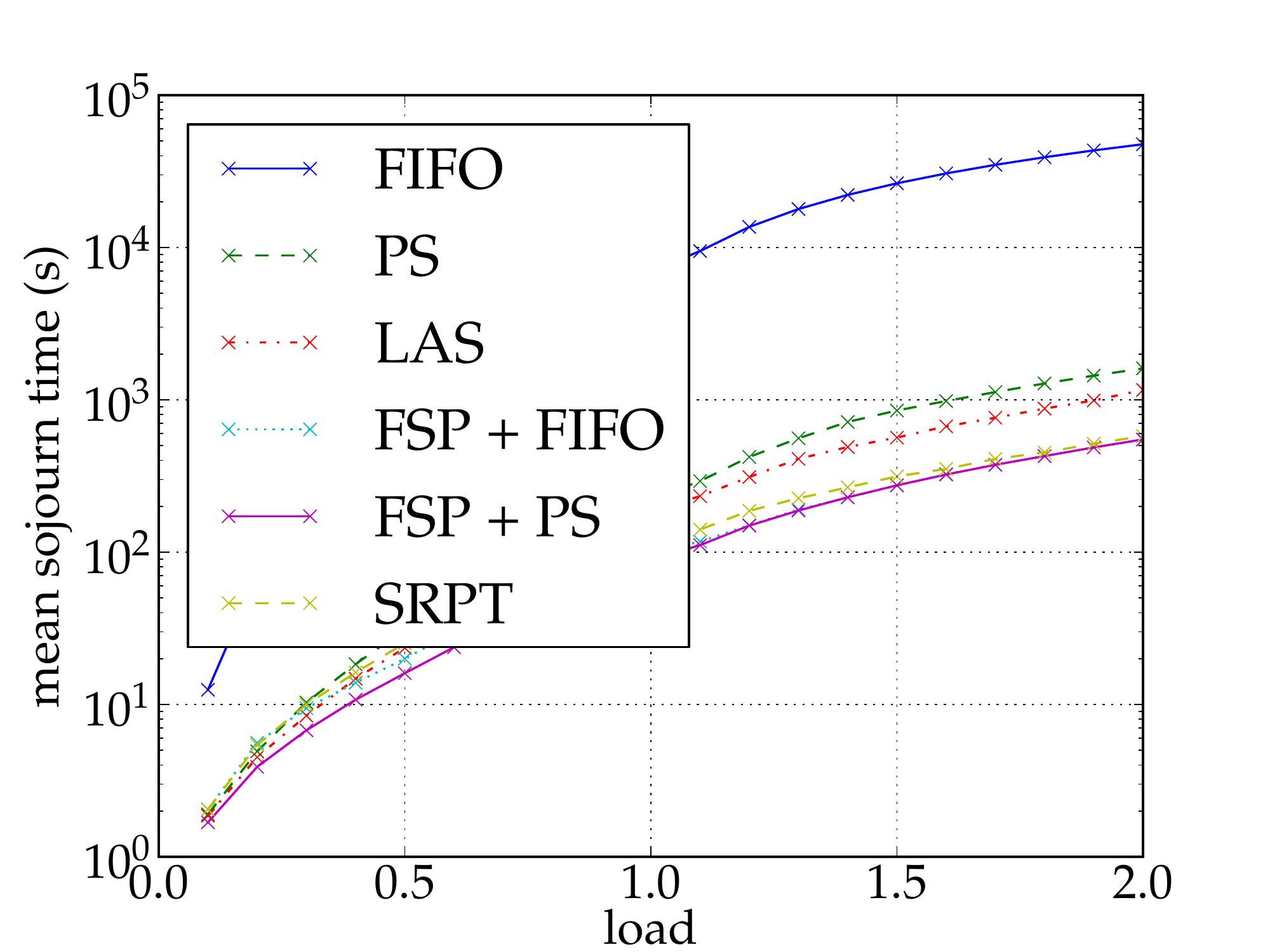}
\par\end{centering}

}\subfloat[\texttt{FB10}.]{\begin{centering}
\includegraphics[width=0.32\linewidth]{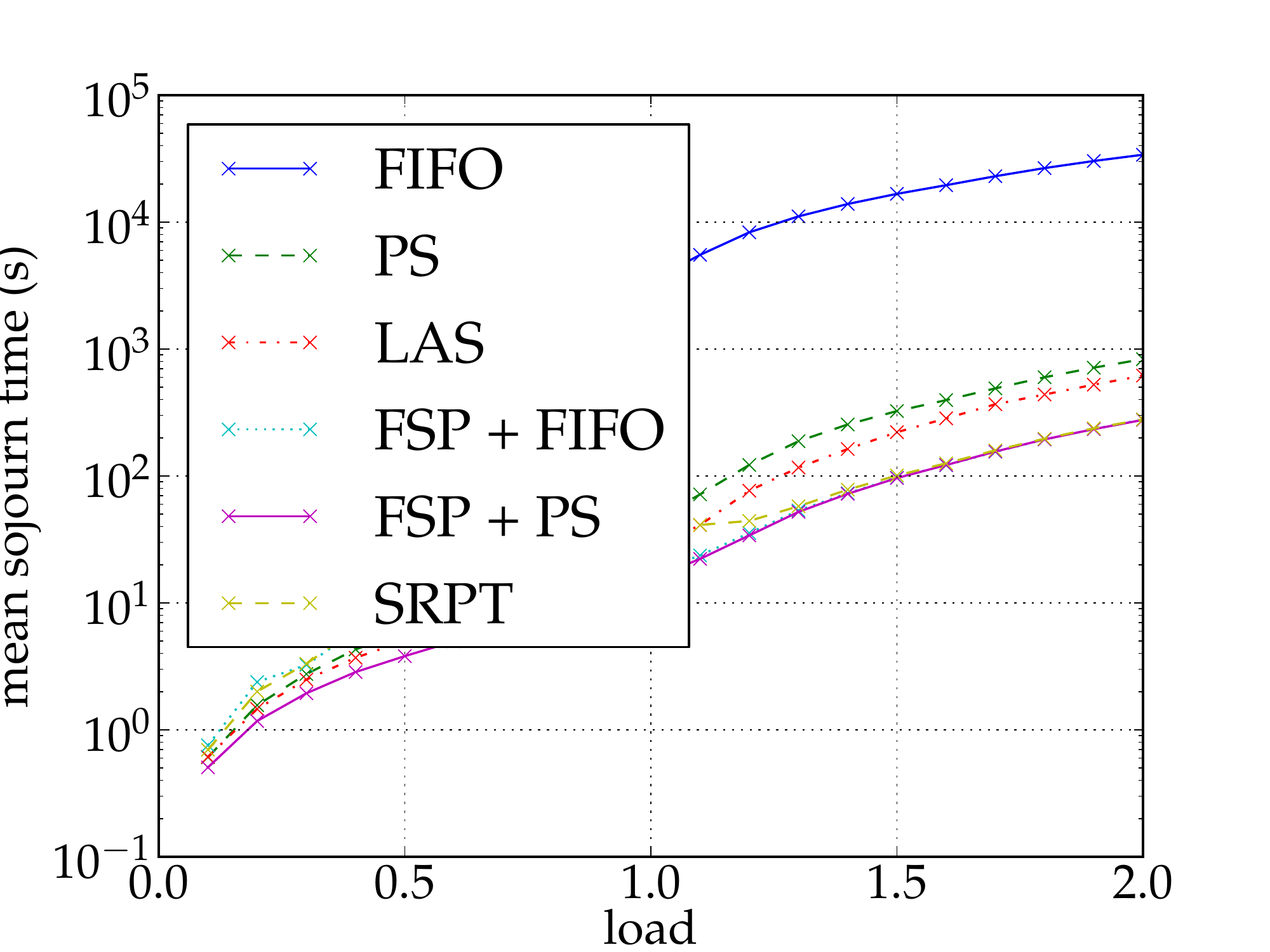}
\par\end{centering}

}\caption{\label{fig:load-0.5}Sojourn versus load: $\sigma=0.5$.}
\end{figure*}

Figure \vref{fig:load-0.5} shows instead the evolution of main sojourn
times for different values of load and $\sigma=0.5$. Obviously, in
this case the results of FIFO and PS do not change: we keep them for
reference. We confirm that, even when varying load, FSP+PS always
performs best. SRPT and FSP+FIFO both suffer from the presence of
error, as we already remarked in Section~\ref{sub:sigma}, but when
load grows beyond 1, differences between algorithms start to become
smaller. The reason for such phenomenon is matter for further study.

\subsection{Sojourn versus $d/n$}

\begin{figure*}
\centering{}\subfloat[\texttt{FB09-0}.]{\begin{centering}
\includegraphics[width=0.32\linewidth]{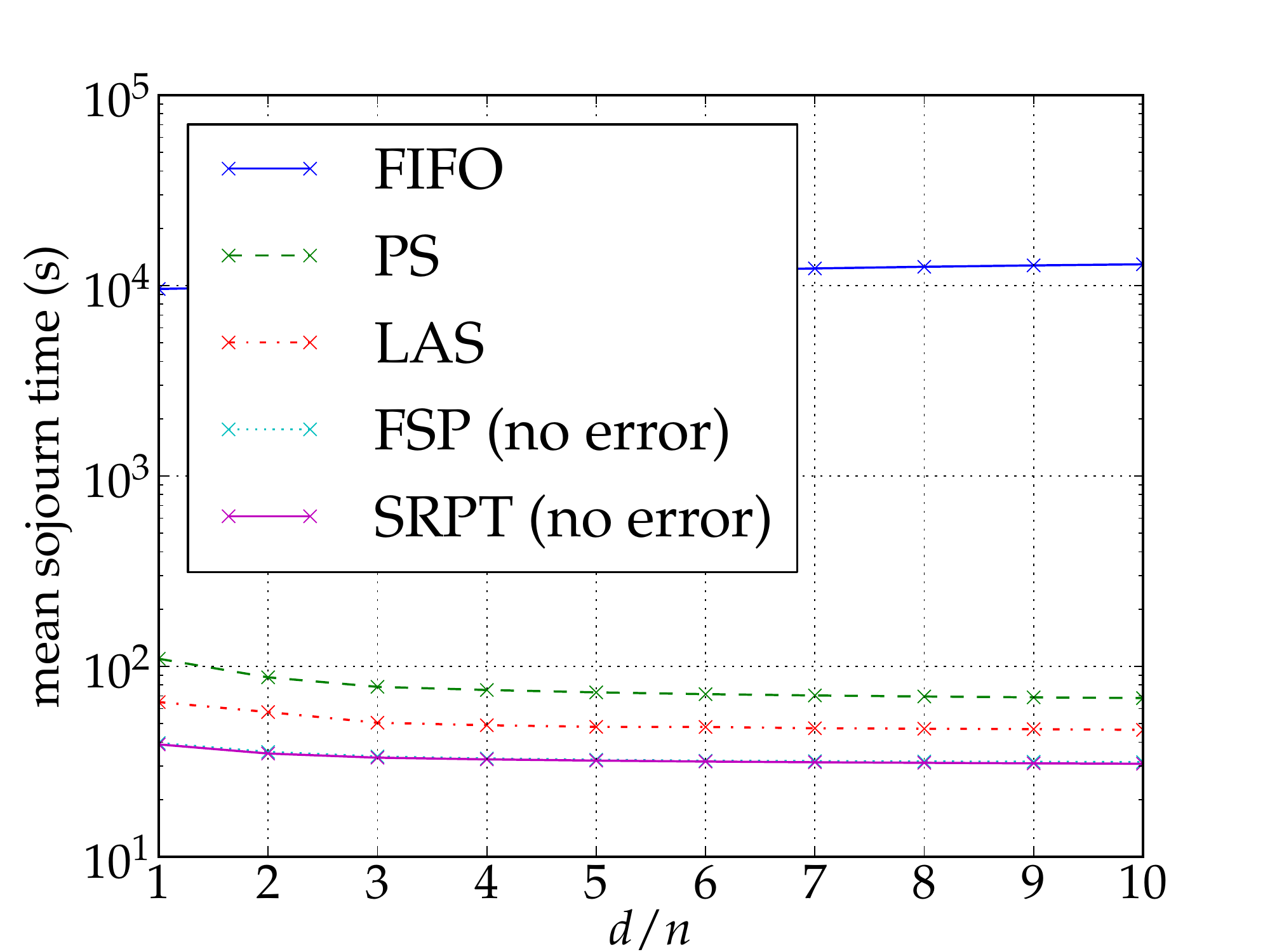}
\par\end{centering}

}\subfloat[\texttt{FB09-1}.]{\begin{centering}
\includegraphics[width=0.32\linewidth]{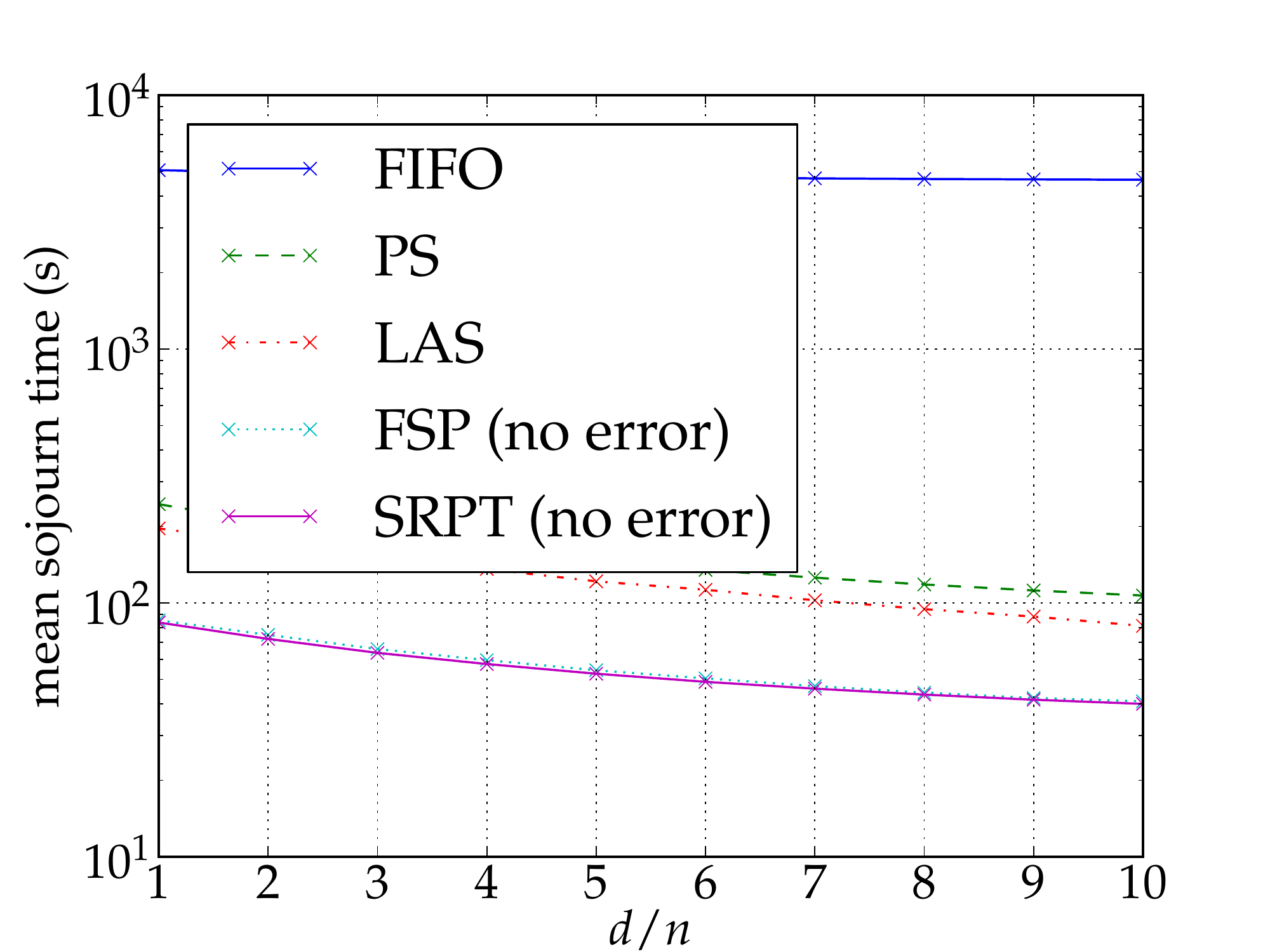}
\par\end{centering}

}\subfloat[\texttt{FB10}.]{\begin{centering}
\includegraphics[width=0.32\linewidth]{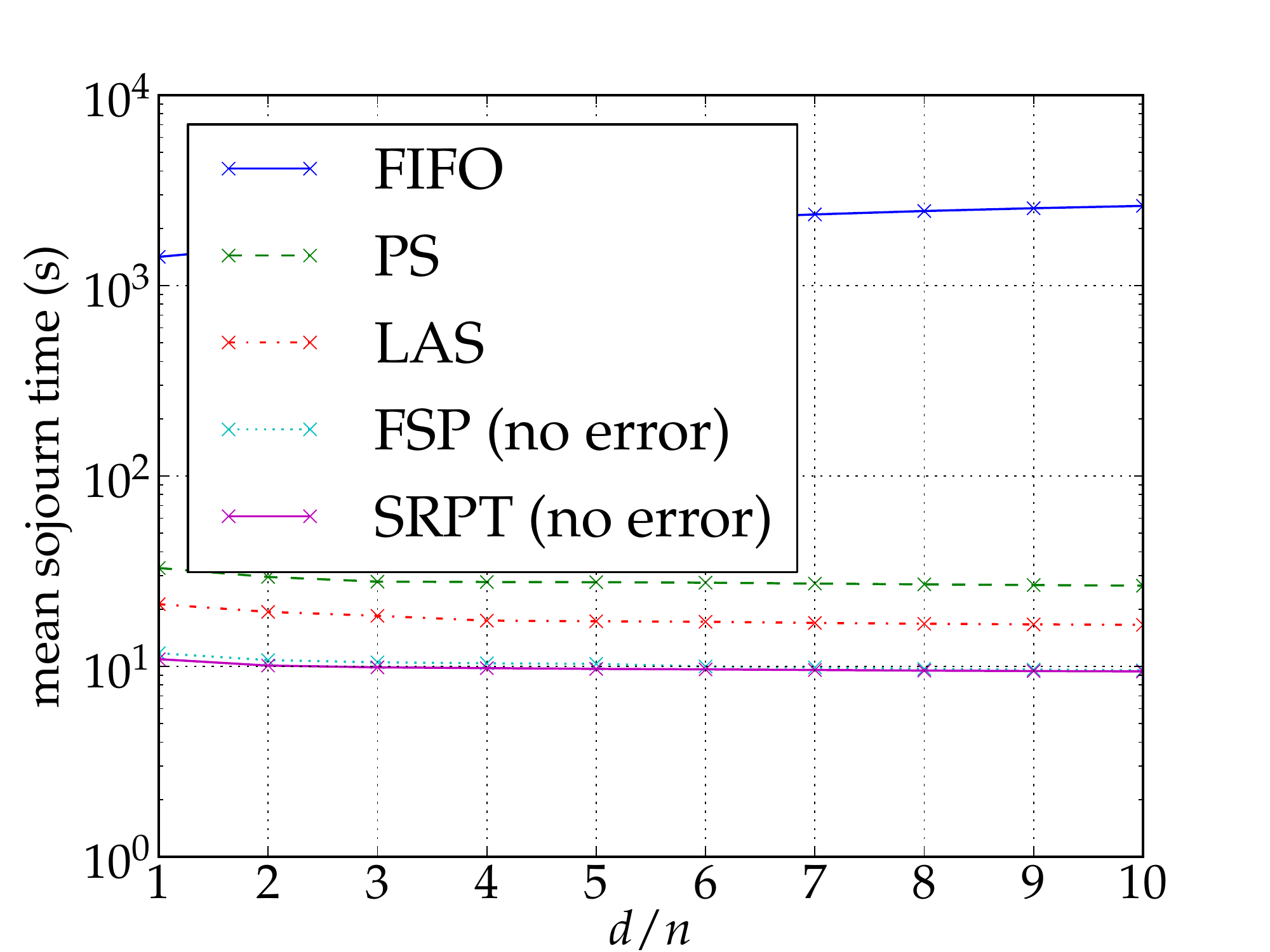}
\par\end{centering}

}\caption{\label{fig:dn-0}Sojourn versus $d/n$: no error.}
\end{figure*}

\begin{figure*}
\centering{}\subfloat[\texttt{FB09-0}.]{\begin{centering}
\includegraphics[width=0.32\linewidth]{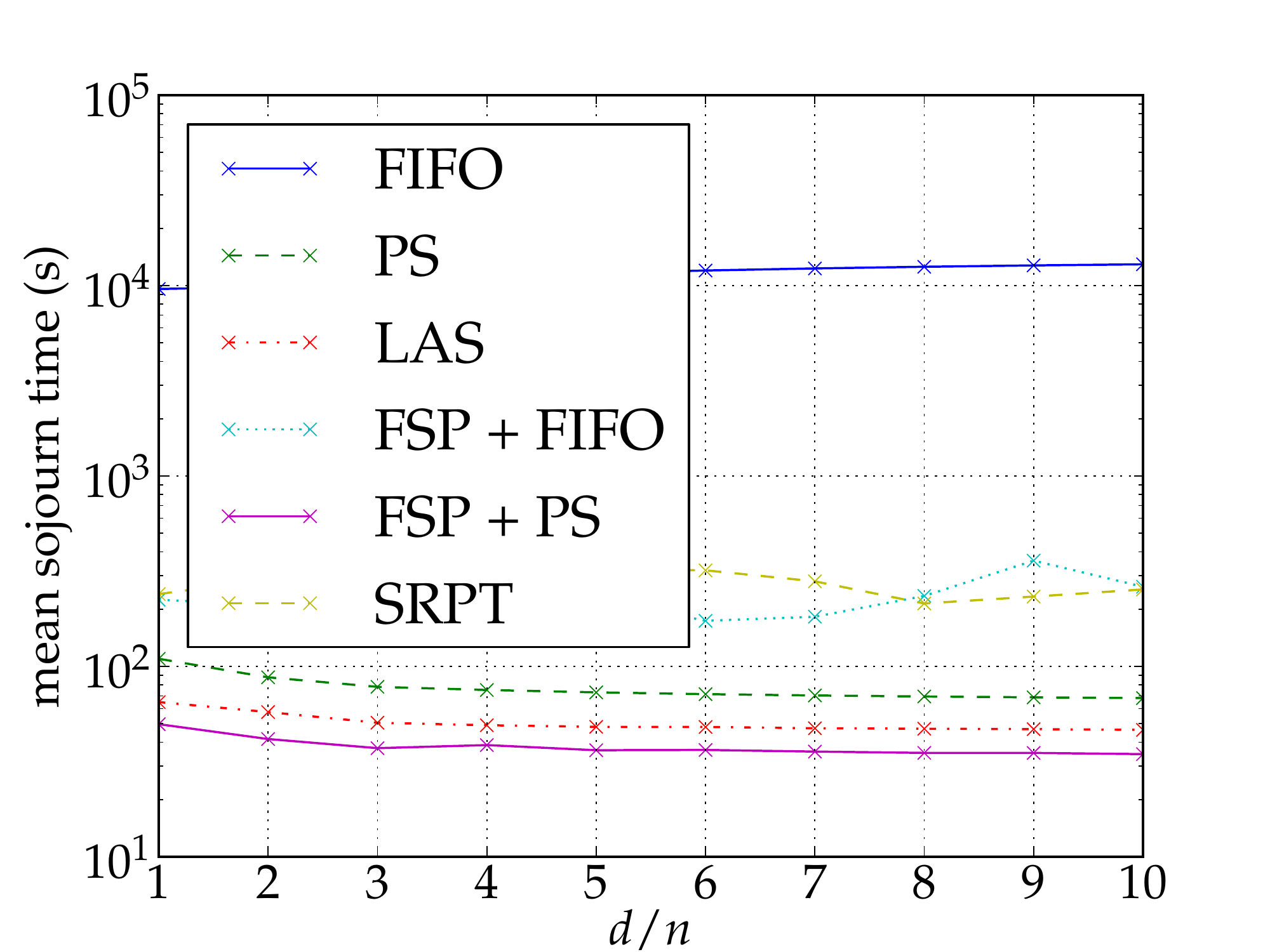}
\par\end{centering}

}\subfloat[\texttt{FB09-1}.]{\begin{centering}
\includegraphics[width=0.32\linewidth]{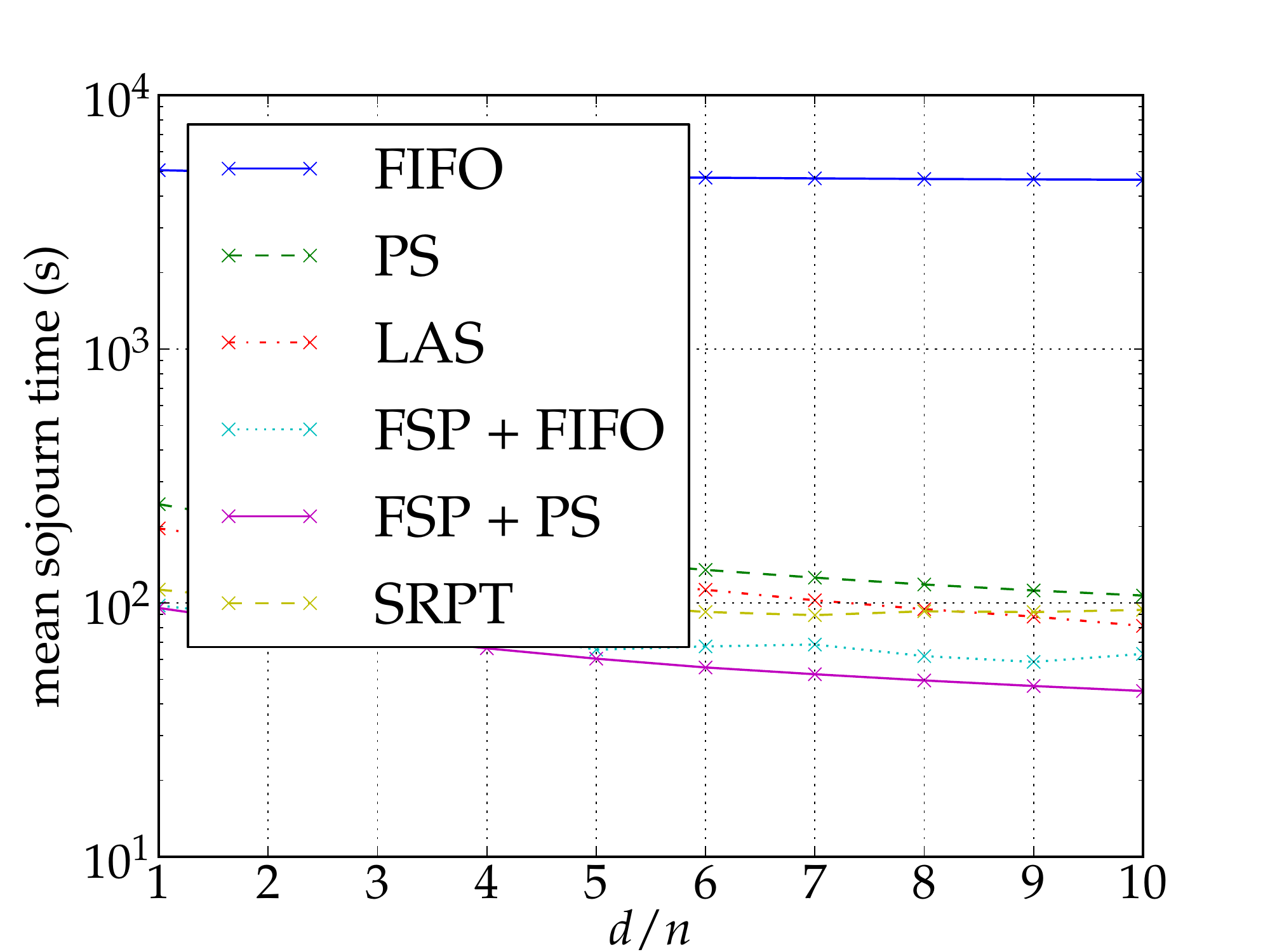}
\par\end{centering}

}\subfloat[\texttt{FB10}.]{\begin{centering}
\includegraphics[width=0.32\linewidth]{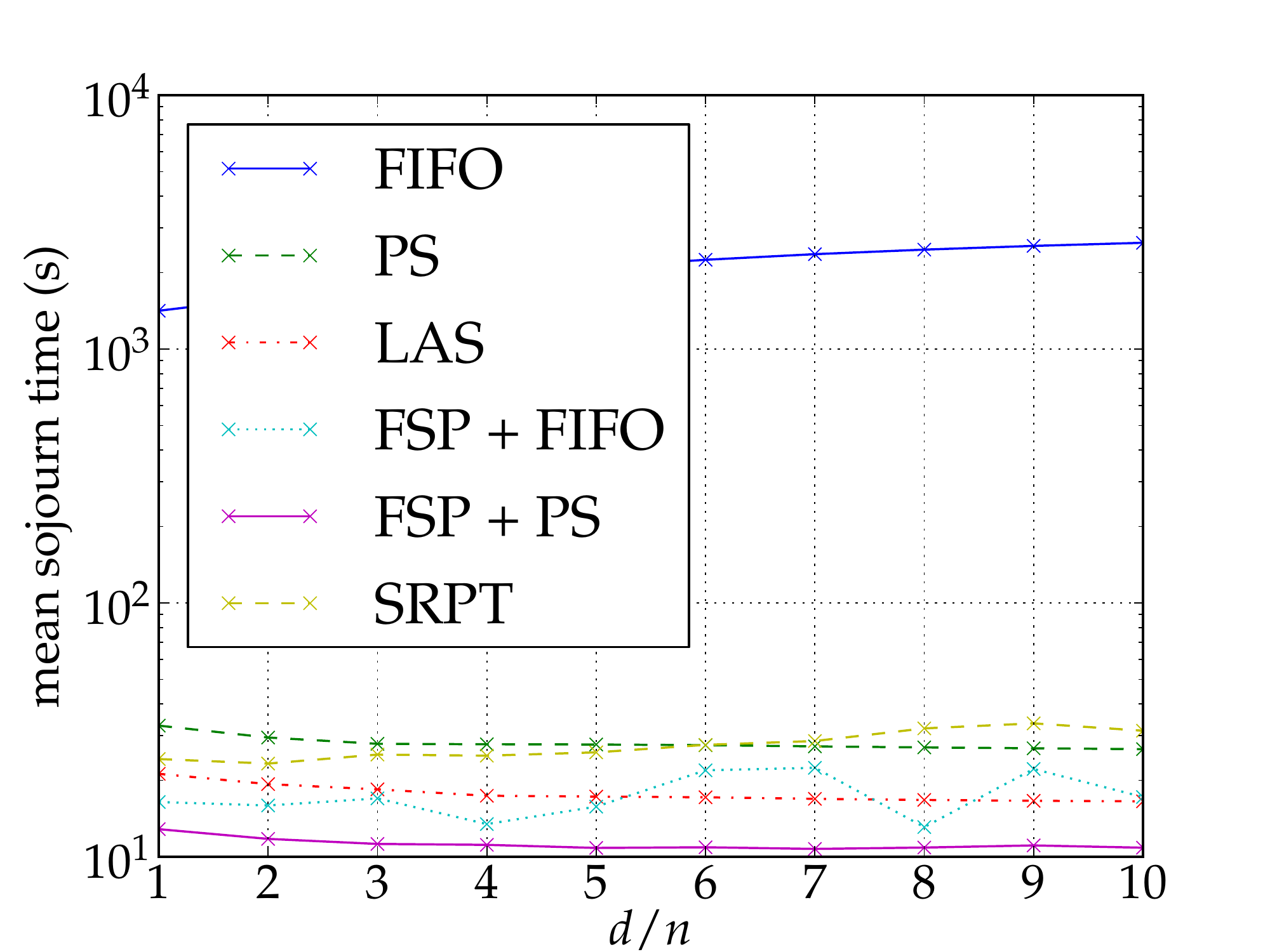}
\par\end{centering}

}\caption{\label{fig:dn-0.5}Sojourn versus $d/n$: $\sigma=0.5$}
\end{figure*}

We conclude our analysis by evaluating the sensitivity of the system
to the $d/n$ parameter. Figures \ref{fig:dn-0} and \vref{fig:dn-0.5}
show that the $d/n$ parameter, required to create the workloads in
our format, doesn't play an important role with respect to scheduling.
We notice, however, that the FSP+FIFO line is much less flat than
the others: the quite random presence of outlier experiments with
very large sojourn times (as already observed in Section~\ref{sub:sigma})
makes the results of this case more noisy.

\section{Conclusions}

This work provides a simulation-based exploration about the applicability
of size-based schedulers in the field of data-intensive computing,
based both on load characteristics from application traces and on
the fact that job size can only be approximated. Our results are very
promising, as they show that size-based scheduling is very beneficial
even when job size can only be approximated very roughly. Our simulator
is available as free software, and we used these simulation results
to help us in the design of the HFSP Hadoop scheduler~\cite{hfsp_arxiv},
which is available as free software as well.%
\footnote{\url{https://bitbucket.org/bigfootproject/hfsp}%
}

We consider this as work in progress, as there are various other points
we are going to explore. To have a better view at the fairness obtained
by the different schedulers, we want to examine \emph{slowdown}, that
is the ratio between a job's size and its sojourn time; we want to
perform a more focused analysis of the three datasets we are currently
examining in order to better understand the difference in terms of
experimental results between them; finally, we want to perform a closer
inspection to the difference in performance between the FSP+PS and
the FSP+FIFO schedulers, in order to obtain a clearer view of their
difference in performance, and investigate whether better solutions
are possible.

\section*{Acknowledgements}

Matteo Dell'Amico is supported by the EU projects BigFoot
(FP7-ICT-223850) and mPlane (FP7-ICT-318627).

\bibliographystyle{IEEEtran}
\bibliography{RR_13_282}

\begin{thebibliography}{10}
\providecommand{\url}[1]{#1}
\csname url@samestyle\endcsname
\providecommand{\newblock}{\relax}
\providecommand{\bibinfo}[2]{#2}
\providecommand{\BIBentrySTDinterwordspacing}{\spaceskip=0pt\relax}
\providecommand{\BIBentryALTinterwordstretchfactor}{4}
\providecommand{\BIBentryALTinterwordspacing}{\spaceskip=\fontdimen2\font plus
\BIBentryALTinterwordstretchfactor\fontdimen3\font minus
  \fontdimen4\font\relax}
\providecommand{\BIBforeignlanguage}[2]{{%
\expandafter\ifx\csname l@#1\endcsname\relax
\typeout{** WARNING: IEEEtran.bst: No hyphenation pattern has been}%
\typeout{** loaded for the language `#1'. Using the pattern for}%
\typeout{** the default language instead.}%
\else
\language=\csname l@#1\endcsname
\fi
#2}}
\providecommand{\BIBdecl}{\relax}
\BIBdecl

\bibitem{schrage1966queue}
L.~E. Schrage and L.~W. Miller, ``The queue m/g/1 with the shortest remaining
  processing time discipline,'' \emph{Operations Research}, vol.~14, no.~4,
  1966.

\bibitem{fsp}
E.~Friedman and S.~Henderson, ``Fairness and efficiency in web server
  protocols,'' in \emph{Proc. of ACM SIGMETRICS}, 2003.

\bibitem{workloads}
Y.~Chen, S.~Alspaugh, and R.~Katz, ``Interactive query processing in big data
  systems: A cross-industry study of mapreduce workloads,'' in \emph{Proc. of
  VLDB}, 2012.

\bibitem{workloads_research}
K.~Ren, Y.~Kwon, M.~Balazinska, and B.~Howe, ``Hadoop's adolescence: A
  comparative workload analysis from three research clusters,'' in
  \emph{Technical Report, CMU-PDL-12-106}, 2012.

\bibitem{ARIA11}
A.~Verma, L.~Cherkasova, and R.~H. Campbell, ``{ARIA: automatic resource
  inference and allocation for mapreduce environments},'' in \emph{Proc. of
  ICAC}, 2011.

\bibitem{mascots12}
------, ``{Two Sides of a Coin: Optimizing the Schedule of MapReduce Jobs to
  Minimize Their Makespan and Improve Cluster Performance},'' in \emph{Proc. of
  IEEE MASCOTS}, 2012.

\bibitem{nsdi12-c}
S.~Agarwal \emph{et~al.}, ``{Re-optimizing Data-Parallel Computing},'' in
  \emph{Proc. of USENIX NSDI}, 2012.

\bibitem{query_perf}
A.~D. Popescu \emph{et~al.}, ``Same queries, different data: Can we predict
  query performance?'' in \emph{Proc. of SMDB}, 2012.

\bibitem{inaccuratesizebasedscheduling}
D.~Lu, H.~Sheng, and P.~Dinda, ``Size-based scheduling policies with inaccurate
  scheduling information,'' in \emph{Proc. of IEEE MASCOTS}, 2004.

\bibitem{hfsp_arxiv}
M.~Pastorelli, A.~Barbuzzi, D.~Carra, M.~Dell'Amico, and P.~Michiardi,
  ``Practical size-based scheduling for {MapReduce} workloads,'' \emph{CoRR},
  vol. abs/1302.2749, 2013.

\bibitem{tinytasks}
K.~Ousterhout, A.~Panda, J.~Rosen, S.~Venkataraman, R.~Xin, S.~Ratnasamy,
  S.~Shenker, and I.~Stoica, ``The case for tiny tasks in compute clusters,''
  in \emph{Proceedings of the 14th USENIX conference on Hot Topics in Operating
  Systems}.\hskip 1em plus 0.5em minus 0.4em\relax USENIX Association, 2013.

\bibitem{flat-datacenter-storage}
E.~B. Nightingale, J.~Elson, J.~Fan, O.~Hofmann, J.~Howell, and Y.~Suzue,
  ``Flat datacenter storage,'' in \emph{Proceedings of the 10th USENIX
  conference on Operating systems design and implementation}, 2012.

\bibitem{swim_tool}
Y.~Chen, S.~Alspaugh, A.~Ganapathi, R.~Griffith, and R.~Katz, ``Statistical
  workload injector for {MapReduce},''
  \url{https://github.com/SWIMProjectUCB/SWIM}.

\bibitem{eurosys10}
M.~Zaharia \emph{et~al.}, ``Delay scheduling: A simple technique for achieving
  locality and fairness in cluster scheduling,'' in \emph{Proc. of ACM
  EuroSys}, 2010.

\bibitem{swim}
Y.~Chen, A.~Ganapathi, R.Griffith, and R.~Katz, ``The case for evaluating
  mapreduce performance using workload suites,'' in \emph{Proc. of IEEE
  MASCOTS}, 2011.

\bibitem{Rai:2003:ALS:885651.781055}
\BIBentryALTinterwordspacing
I.~A. Rai, G.~Urvoy-Keller, and E.~W. Biersack, ``Analysis of las scheduling
  for job size distributions with high variance,'' \emph{SIGMETRICS Perform.
  Eval. Rev.}, vol.~31, no.~1, pp. 218--228, Jun. 2003. [Online]. Available:
  \url{http://doi.acm.org/10.1145/885651.781055}
\BIBentrySTDinterwordspacing

\bibitem{moseley_soda}
K.~Fox and B.~Moseley, ``Online scheduling on identical machines using
  {SRPT},'' in \emph{In Proc. of ACM-SIAM SODA}, 2011.

\bibitem{fair_queuing}
J.~Nagle, ``On packet switches with infinite storage,'' \emph{Communications,
  IEEE Transactions on}, vol.~35, no.~4, 1987.

\bibitem{gorinsky2007fair}
S.~Gorinsky and C.~Jechlitschek, ``Fair efficiency, or low average delay
  without starvation,'' in \emph{Proc. of IEEE ICCCN}, 2007.

\end{thebibliography}

\end{document}